\documentstyle[aps,epsfig,multicol]{revtex}

\title{ Detection of Single Spin Decoherence in a Quantum Dot via Charge Currents
}

\author{  Hans-Andreas Engel\cite{EmailEngel}
  and Daniel Loss\cite{EmailLoss}
}

\address{Department of Physics and Astronomy, University of Basel,
 Klingelbergstrasse 82, CH-4056 Basel, Switzerland}

\newcommand{\etal}[1]{ {\it et al.}}

\newcommand{\figwidth}{\ifpreprintsty400pt\else8.6cm\fi}

\newcommand{\mubohr}{{\mu_{\rm B}}}
\newcommand{\efermi}{{\varepsilon_{\rm F}}}

\newcommand{\bra}[1]{{\langle #1 |}}
\newcommand{\ket}[1]{{| #1 \rangle}}

\newcommand{\expect}[1]{{\left\langle #1 \right\rangle}}

\newcommand{\spup}{\ket{\!\uparrow}}
\newcommand{\spdown}{\ket{\!\downarrow}}

\newcommand{\spupdown}{\ket{\!\uparrow\downarrow}}
\newcommand{\spdownup}{\ket{\!\downarrow\uparrow}}

\newcommand{\du}{{\downarrow\uparrow}}
\newcommand{\ud}{{\uparrow\downarrow}}

\newcommand{\rhoElem}[1]{{\rho_{#1}}}
\newcommand{\rhoDotElem}[1]{{\dot{\rho}_{#1}}}
\newcommand{\rhouu}{{\rhoElem{\uparrow}}}
\newcommand{\rhodd}{{\rhoElem{\downarrow}}}
\newcommand{\rhoSS}{{\rhoElem{S}}}

\newcommand{\rhodu}{{\rhoElem{\du}}}

\newcommand{\rhoSu}{{\rhoElem{S \uparrow}}}

\newcommand{\rhoSd}{{\rhoElem{S \downarrow}}}

\newcommand{\rhoDotuu}{{\rhoDotElem{\uparrow}}}
\newcommand{\rhoDotdd}{{\rhoDotElem{\downarrow}}}
\newcommand{\rhoDotSS}{{\rhoDotElem{S}}}

\newcommand{\rhoDotdu}{{\rhoDotElem{\du}}}

\newcommand{\rhoDotSu}{{\rhoDotElem{S \uparrow}}}

\newcommand{\rhoDotSd}{{\rhoDotElem{S \downarrow}}}

\newcommand{\gammaP}{V}   
\newcommand{\WP}{W}   
\newcommand{\gammadu}{\gammaP_{\du}}
\newcommand{\gammaSu}{\gammaP_{S \uparrow}}
\newcommand{\gammaSd}{\gammaP_{S \downarrow}}
\newcommand{\WdS}{{\WP_{\downarrow S}}}
\newcommand{\WSd}{{\WP_{S \downarrow}}}
\newcommand{\WuS}{{\WP_{\uparrow S}}}
\newcommand{\WSu}{{\WP_{S \uparrow}}}
\newcommand{\WdSl}[1]{{\WP_{\downarrow S}^{#1}}}
\newcommand{\WSdl}[1]{{\WP_{S \downarrow}^{#1}}}
\newcommand{\WuSl}[1]{{\WP_{\uparrow S}^{#1}}}
\newcommand{\WSul}[1]{{\WP_{S \uparrow}^{#1}}}

\newcommand{\Wud}{\WP_{\ud}}
\newcommand{\Wdu}{\WP_{\du}}
\newcommand{\WRud}{{\widetilde{\WP}_{\ud}}}
\newcommand{\WRdu}{{\widetilde{\WP}_{\du}}}
\newcommand{\Womegarf}{\WP_{\omegarf}}
\newcommand{\WoMax}{\WP_{\omegarf}^{\rm max}}

\newcommand{\ES}{E_{S}}
\newcommand{\ETp}{E_{T_+}}
\newcommand{\ESu}{\Delta_{S\uparrow}}
\newcommand{\ESd}{\Delta_{S\downarrow}}

\newcommand{\gammaTZero}{\gamma}

\newcommand{\gul}[1]{\gammaTZero^\uparrow_{#1}}
\newcommand{\gdl}[1]{\gammaTZero^\downarrow_{#1}}
\newcommand{\guO}{\gul{1}}
\newcommand{\guT}{\gul{2}}
\newcommand{\gdO}{\gdl{1}}
\newcommand{\gdT}{\gdl{2}}
\newcommand{\gO}{\gammaTZero_1}
\newcommand{\gT}{\gammaTZero_2}
\newcommand{\tmeas}{t_{\rm meas}}
\newcommand{\hgu}{\hat{\gammaTZero}^{\uparrow}}
\newcommand{\hgd}{\hat{\gammaTZero}^{\downarrow}}

\newcommand{\Hdot}{{H_{\rm dot}}}
\newcommand{\HRdot}{{\tilde{H}_{\rm dot}}}
\newcommand{\Hrf}{{H_{\rm ESR}}}

\newcommand{\HrfOsc}{{\cos{(\omegarf  t)}}}
  
\newcommand{\HrfConst}{{\Delta_x}} 

\newcommand{\omegarf}{{\omega}}
\newcommand{\HrfOffDiag}{{\HrfConst \HrfOsc}}

\newcommand{\HDD}{H_{\rm DD}}
\newcommand{\HDL}{H_{\rm DL_2}}
\newcommand{\tDD}{t_{\rm DD}}
\newcommand{\tDL}{t_{\rm DL_2}}

\newcommand{\qdot}[1]{\begin{picture}(14,11)
    \put(6,3.6){\circle{13}}
    \put(6,3.6){\makebox(0,0){$#1$}}
    \end{picture}}
\newcommand{\edot}[1]{\begin{picture}(7,9) 
    \put(3,3.6){\makebox(0,0){$#1$}}
    \end{picture}}
\newcommand{\qddot}[2]{\qdot{#1}$_{\!1}$\qdot{#2}$_{\!2}$}

\begin{document}

\maketitle

\begin{abstract}
We consider a quantum dot attached to leads in the
Coulomb blockade regime which has a spin 1/2
ground state. We show that by applying an ESR field to the
dot-spin the stationary current in the sequential tunneling regime
exhibits a resonance whose line width is determined by
the single-spin decoherence time $T_2$. 
The Rabi oscillations of the dot-spin are shown to induce
coherent current oscillations from which $T_2$ can be deduced
in the time domain. We describe a spin-inverter which
can be used to pump current through a double-dot via spin flips
generated by ESR.
\end{abstract}


\ifpreprintsty\else\begin{multicols}{2}\fi

An increasing number of spin-related
 experiments~\cite{Kikkawa,Gupta,Roukes,Fiederling,Ensslin,Fujisawa}
 show that the
 electron spin is a robust candidate
 for coherent quantum state engineering
 in solid state systems such as semiconductor nanostructures.
Several techniques, most prominently 
 electron spin resonance (ESR), can then be envisaged
 for  manipulation 
 of electron spins on quantum dots \cite{Loss97,QCReview},
where the coherence of the spin is limited by 
 the intrinsic
 spin decoherence time $T_2$.
In some related systems,
 time-resolved optical measurements were used to determine $T_2^{*}$,
 the decoherence time
 of many of spins, with $T_2^{*}$
 exceeding 100 ns in bulk GaAs \cite{Kikkawa}.
More recently,
 the single spin relaxation time $T_1$
 (generally $T_1 \geq T_2$)
 of a single quantum dot attached to leads
 was measured via transport to be longer than a few $\mu$s \cite{Fujisawa},
 consistent with calculations \cite{KhaetskiiNazarov}.
In this work, we go one step further and propose a setup
to
 extract the single spin decoherence time $T_2$
 of an electron confined in a quantum dot from transport
measurements.
 The dot, which is
 attached to leads, is operated in the Coulomb blockade regime,
 and the spin flips generated by an ESR source
 lead to a resonance in the stationary charge current
with a line width determined by the spin decoherence time, see Figs. 1,2.
Making use of coupled master equations we analyze the
time-dependence of the current and the spin-measurement
process (read-out)\cite{Schoen},
and show that coherent Rabi oscillations of the dot-spin induce
 oscillations of the current, providing a measure of
the spin decoherence directly in time space, see Fig. 3.
In the absence of a bias, the current can be pumped through a double-dot
with the ESR source (providing the necessary energy via spin flips on the dot)
and by making use of
a novel spin-inverter for producing spin-dependent tunneling.

{\it Model.}
We study a quantum dot in Coulomb blockade regime \cite{kouwenhoven},
 coupled to two Fermi-liquid leads $l=1$, $2$
 at chemical potentials $\mu_l$.
We consider the Hamiltonian $H = H_0 + H_T = H_{\rm lead} + \HRdot + H_T$,
 which describes leads, dot
 and the tunnel coupling between leads and dot, {\it resp}.
Here, $\HRdot=\Hdot+\Hrf$,
 where $\Hdot$ includes charging and interaction energies of the electrons
 on the dot.
$\Hdot$ also contains a Zeeman coupling term to a constant magnetic field $B_z$
 in $z$-direction,
 $-\frac{1}{2} \Delta_z \sigma_z$, 
 with Zeeman splitting $\Delta_z=g\mubohr B_z$, 
 electron $g$ factor $g$,
 Bohr magneton $\mubohr$, and Pauli matrix $\sigma_z$.
Coupling to an oscillating magnetic ESR field in $x$-direction
 of frequency $\omegarf$
 is included in $\Hrf = - \frac{1}{2} \HrfOffDiag \,\sigma_x$,
 with $\HrfConst = g\mubohr B_x^0$  and Pauli matrix $\sigma_x$.
Such an oscillating field produces Rabi spin-flips at $\omegarf=\Delta_z$,
 as used in ESR.
We assume Zeeman splitting of the leads
 $\Delta_z^{\rm leads}\not\approx\Delta_z$ and
 $\Delta_z^{\rm leads}\ll \efermi$, where $\efermi$ is the Fermi energy,
 such that the field effects of $B_z$ and $B_x(t)$, {\it resp.},
 on the leads are negligible. 
Such a situation can be achieved by using 
 materials of different $g$ factors \cite{Fiederling}
 and/or with local magnetic fields.

\begin{figure}
\centerline{\psfig{file=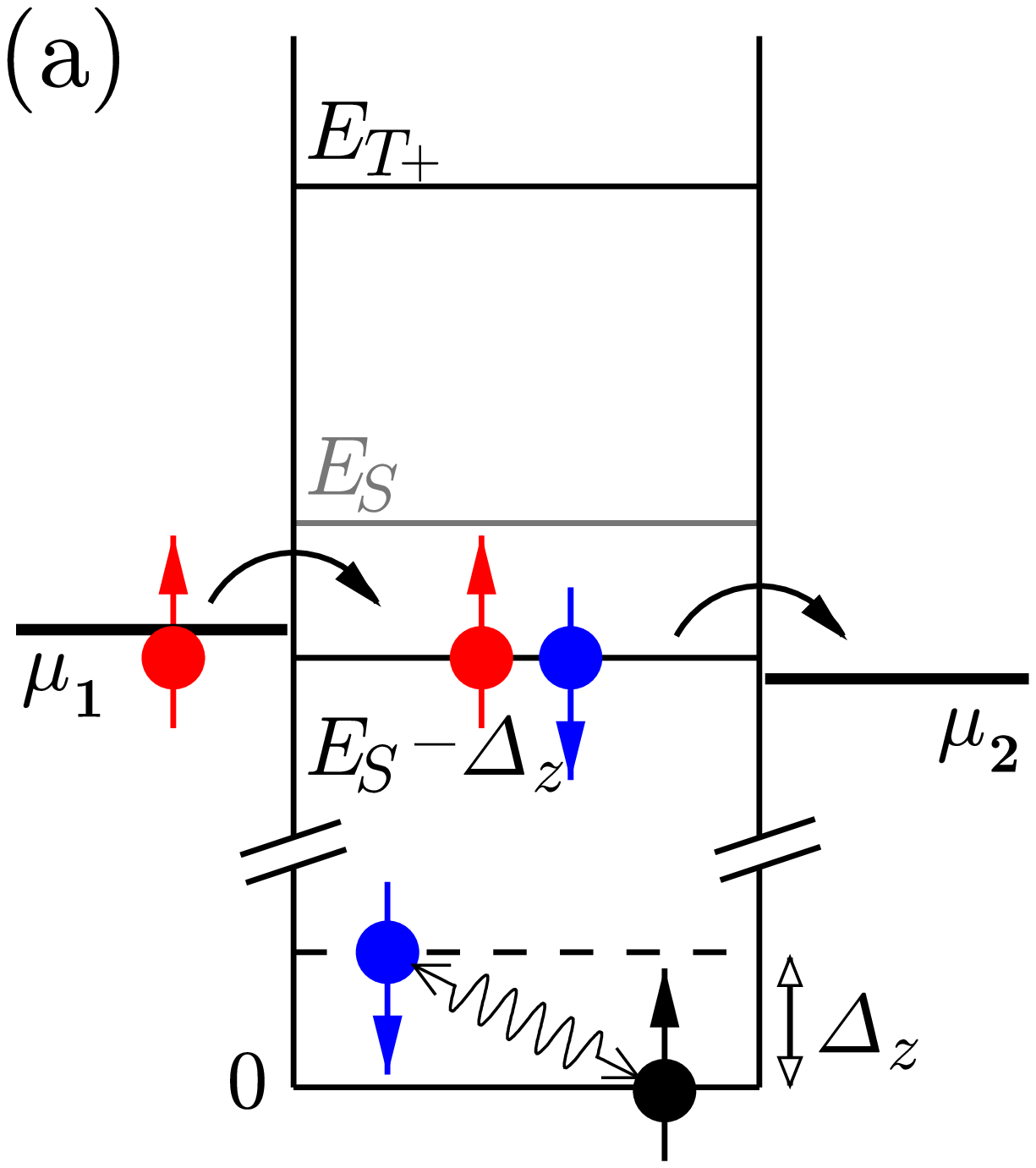,height=3.5cm}\hspace{3mm}\psfig{file=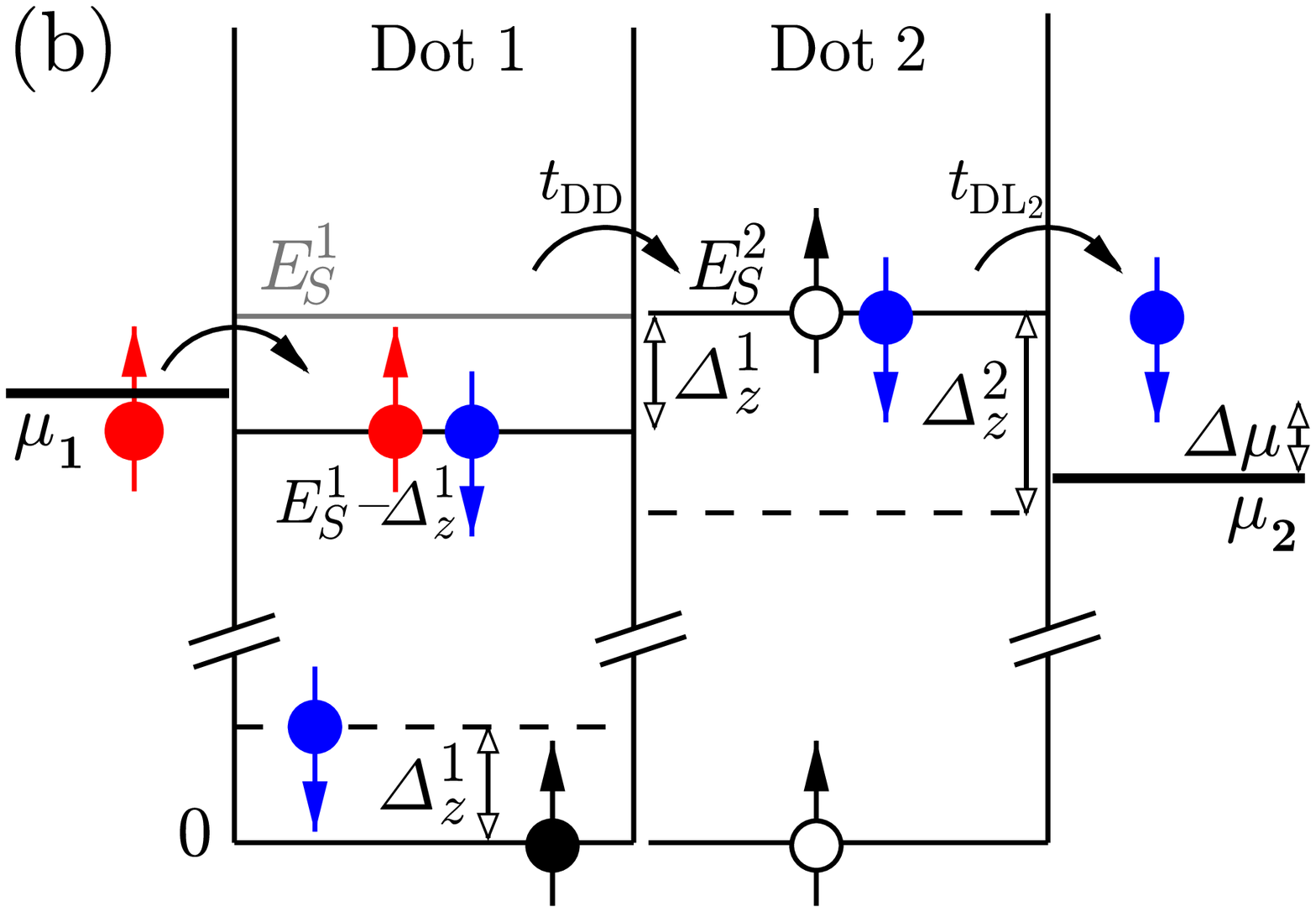,height=3.5cm}}
\caption[Dot]{
\label{figAcDot}
(a) Dot coupled to unpolarized leads with chemical potentials $\mu_{1,\,2}$,
 in the sequential tunneling regime defined by
 $\ES > \mu_1 > \ES-\Delta_z > \mu_2$,
 with the singlet/triplet levels $\ES$/$\ETp$
 and Zeeman splitting $\Delta_z=g\mubohr B_z$.
 If initially the spin-state on the dot is $\spup$, sequential tunneling
 is blocked by energy conservation.
Exciting the dot-spin via ESR (Rabi flip)
 the dot becomes unblocked but only for spin up electrons from lead 1.
Finally, from the singlet, spin up or down can tunnel into lead 2.
(b) Extended setup where the additional dot 2 
 (with $|\tDD|<|\tDL|$ and tuned to resonance)
 acts as a spin filter in the regime
 $\ES^1\approx\ES^2$;
 $\ES^1 > \mu_1 > \ES^1-\Delta_z^1$;
 $\mu_2 > \ES^1-\Delta_z^2$;
 $\ES^2>\mu_2$, and
 $\Delta_z^1\not\approx\Delta_z^2$ \cite{ZeemanDotTwo}.
The allowed transition sequence is schematically given by
 \edot{\uparrow}\qddot{\uparrow}{\uparrow}  
 $\stackrel{\mbox{\tiny ESR}}{\longrightarrow}$
 \edot{\uparrow}\qddot{\downarrow}{\uparrow} $\to$
                \qddot{\ud}{\uparrow}        $\to$
                \qddot{\uparrow}{\ud}      $\leftrightarrow$
                \qddot{\uparrow}{\uparrow}\edot{\downarrow} (see text).
}
\end{figure}

We now describe the electronic states of the dot.
For an odd number of electrons on the dot with antiferromagnetic filling,
 the topmost (excess) electron can
 be either in the spin ground state, $\spup$,
 or in the excited state, $\spdown$, see Fig. \ref{figAcDot}.
For an additional electron on the dot,
 we assume the ground state
 to be the singlet $\ket{S} = (\spupdown-\spdownup)/\sqrt{2}$
 (which can be achieved by tuning $B_z$ \cite{taruchaKouwenhoven}).
The energy of the dot is defined by $\Hdot \ket{n} = E_n \ket{n}$.
It is convenient to use only one-particle energies 
 $\Delta_{S\uparrow(\downarrow)}=E_S-E_{S\uparrow(\downarrow)}$
 (containing charging energy $U$),
 which can then be compared with $\mu_l$.

{\it Master equation}.
We derive
 the master equation for
 the reduced density matrix of the dot, $\rho_D = {\rm Tr}_L\, \rho$.
Here, ${\rm Tr}_L$ is the trace taken over the leads
 and $\rho$ is the full density matrix.
Using a superoperator formalism, 
 we evaluate the von Neumann equation
 within Born approximation in $H_T$ 
 while taking $\Hrf$ fully into account.
Hereby we make the usual assumption
 that the correlations induced in the leads by $H_T$
 decay rapidly (Markovian approximation)
 and that we can neglect non-secular terms \cite{Blum,Details}.
We obtain the following master equation

\begin{eqnarray}
\rhoDotuu &=&
   - (\Wdu+\WSu)\, \rhouu
   + \Wud \,\rhodd
   + \WuS \,\rhoSS
\nonumber \\ & & + \HrfConst\,\HrfOsc \:{\rm Im}[\rhodu],
\label{eqnRhoDotuu}
\\
\rhoDotdd &=&
    \Wdu \,\rhouu
  - (\Wud+\WSd)\,\rhodd
  + \WdS \,\rhoSS
\nonumber \\ & & - \HrfConst\,\HrfOsc \:{\rm Im}[\rhodu],
\label{eqnRhoDotdd}
\\
\rhoDotSS &=&
   \WSu\,\rhouu +\WSd\,\rhodd
  -(\WuS+\WdS)\,\rhoSS   ,
\label{eqnRhoDotSS}
\\
\rhoDotdu &=&
  - i \Delta_z \rhodu
  - i \HrfConst\HrfOsc (\rhouu -\rhodd)/2
 - \gammadu \, \rhodu,\!\!
\label{eqnRhoDotdu}
\\
\rhoDotSu &=&
  - i \ESu \rhoSu - \gammaSu \, \rhoSu  ,
\label{eqnRhoDotSu}
\\
\rhoDotSd &=&
  - i \ESd \rhoSd -  \gammaSd \, \rhoSd  ,
\label{eqnRhoDotSd}
\end{eqnarray}
where $\rho_n = \bra{n}\rho_D\ket{n}$ and
 $\rho_{nm} = \bra{n}\rho_D\ket{m}$.
We now specify the transition rates $\WP_{nm}$ for the diagonal
and the spin decoherence rates  $\gammaP_{nm}=\gammaP_{mn}$
 for the off-diagonal
 elements of $\rho_D$.
For the sequential tunneling rates \cite{CT},
 $\WP_{nm}=\sum_{l=1,2}\WP_{nm}^{l}$,
we find,
 $\WSdl{l}=\gul{l} \, f_l(\ESd)$ and
 $\WdSl{l}=\gul{l} \, [1-f_l(\ESd)]$,
 with the Fermi function $f_l(\ESd)=\big[1+e^{(\ESd-\mu_l)/kT}\big]^{-1}$.
The transition rates are
 $\gul{l} = 2\pi \nu_{\uparrow} | \sum_p t_{lp}
  \bra{\downarrow} d_{p\uparrow} \ket{S}|^2$
 with (possibly spin-dependent, see below) 
 density of states $\nu_{\uparrow}$ 
 at the Fermi energy.
The rates  $\WSul{l}$ and  $\WuSl{l}$ are defined analogously.
Further, we allow for
 additional coupling of the electron spin
 to the environment (e.g.\ hyperfine or spin-phonon coupling).
First, the spin relaxation rates $\Wud$ and $\Wdu$
 were inserted in Eqs.\ (\ref{eqnRhoDotuu}) and (\ref{eqnRhoDotdd}),
 corresponding to the phenomenological rate $1/T_1 = \Wud + \Wdu$ \cite{Blum}.
We assume $\Wud\gg\Wdu$ for $\Delta_z>kT$
 (consistent with detailed balance, $\Wud/\Wdu = e^{\Delta_z/kT}$).
Second, the rate $1/T_2$ describes
 the intrinsic decoherence of the spin on the dot
 (which persists even if the tunnel coupling is switched off),
 contributing to $\gammadu$.
The contribution of $H_T$ to $\gammadu$
 is calculated as $(\WSu+\WSd)/2$,
 i.e.\ electrons tunneling onto the dot destroy spin coherence
 on the dot.
The total spin decoherence rate is
  $\gammadu = (\WSu+\WSd)/2 + 1/T_2$.

We calculate the stationary solution of 
 Eqs.\ (\ref{eqnRhoDotuu})--(\ref{eqnRhoDotSd})
 in the rotating wave approximation \cite{Blum},
 where only the leading frequency contributions
 of $\Hrf$ are retained.
We obtain effective spin-flip rates
 $\WRud = \Wud+\Womegarf$ and $\WRdu = \Wdu+\Womegarf$,
 where the Rabi-flips produced by the ESR field are described
 by the rate
\begin{eqnarray}
\label{eqnWomegarf}
\Womegarf   
= \frac{\HrfConst^2}{8}
   \frac{\gammadu}{(\omegarf - \Delta_z)^2 + \gammadu^2},
\end{eqnarray}
 which is a Lorentzian in $\omega$
 with maximum $\WoMax=\HrfConst^2/8\gammadu$
 at ESR resonance $\omegarf=\Delta_z$.
With the stationary solution, we can now calculate the 
 current $I$ (too lengthy to be shown here \cite{Details}),
 which we shall discuss next in different regimes.

\begin{figure} 
\centerline{\psfig{file=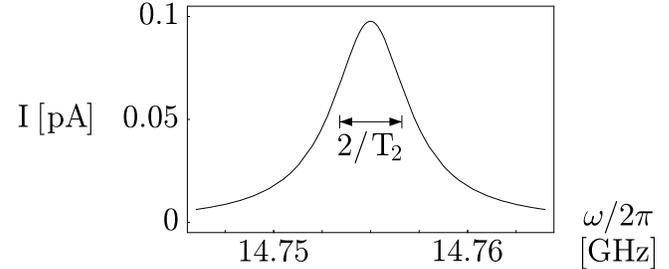,width=\figwidth}}
\caption[Current]{
\label{figCurrent}
The stationary current $I(\omegarf)$ [Eq.\ (\ref{eqnCurrentZeroT})]
 for $B_z = 0.5\: {\rm T}$,
 $B_x^0=0.45\: {\rm G}$,
 $T_1 = 1  \: \mu{\rm s}$,
 $T_2 = 100\: {\rm ns}$,
 $\gO=5\times10^6 \: {\rm s}^{-1}$, and
 $\gT=5\gO$, i.e.\ $\WoMax<\gO<1/T_2$.
Here, the linewidth gives a lower bound for
 the intrinsic spin decoherence time $T_2$,
 while it becomes equal to $2/T_2$ for 
 $B_x^0=0.08\: {\rm G}$ and
 $\gO=5\times10^5 \: {\rm s}^{-1}$, where
 $I(\omegarf=\Delta_z) \approx 1.5\: {\rm fA}$.
}
\end{figure}

{\it Zeeman blockade.}
We consider a quantum dot as shown in Fig.~\ref{figAcDot},
 with $\Delta_z>\Delta\mu,\, kT$, and
 $\ESu>\mu_1>\ESd>\mu_2$, and $f_l(\ESu)=0$,
 with $\Delta\mu=\mu_1-\mu_2$ being the applied bias.
Thus, $\WSu=0$ and $\WuSl{l}=\gdl{l}$.
Without ESR field,
 the dot relaxes into its ground state $\spup$ (since $\Wdu\ll\Wud$),
 and the sequential tunneling 
 current through the dot is blocked by energy conservation.
However, if an ESR field is present, 
 producing Rabi spin-flips (on the dot only), 
 the current flows through the dot involving state $\spdown$.
First we consider $\gdl{l}=\gul{l} = \gammaTZero_l$.
For $kT>\Delta\mu$ and
 $\WoMax < \mbox{max} \{\Wud,\, \gO \}$
 we obtain for the stationary current
\begin{eqnarray}
\label{eqnCurrentZeroT}
I(\omega) &=&
\frac{ 2 e \, \gO \gT \WRdu }{
    \Wud  (\gO \!+\! 2\gT )
+ \gO (\gO \!+\! \gT)   },
\end{eqnarray}
while, for 
 $\WoMax < \mbox{max} \{\Wud,\, \gO f_1(\ESd+\Delta\mu/2) \}$, 
\begin{eqnarray}
\label{eqnCurrentLinResp}
I(\omega) &=&
 \frac{e \, \gO \gT \WRdu}{\gO+\gT}\,
 \,\frac{\Delta\mu}{2 kT  h(T)} 
 \,\cosh^{-2}\!\left( \frac{\ESd-\mu}{2kT} \right),
\end{eqnarray}
 for $kT>\Delta\mu$ and with $\mu=(\mu_1+\mu_2)/2$.
The standard sequential-tunneling peak-shape in Eq.\ (\ref{eqnCurrentLinResp})
 is modified by
$h(T) = 2 \Wud + 
 (\gO\!+\!\gT\!-\!\Wud)\, f_1(\ESd+\Delta\mu/2)$,
 which can affect position and width of the peak.
Most important, the current $I(\omegarf)$ 
 [Eqs.\ (\ref{eqnCurrentZeroT}) and (\ref{eqnCurrentLinResp})]
 is proportional to the resonant rate $\Womegarf$.
Thus, the current $I(\omegarf)$ 
 as a function of the ESR frequency $\omegarf$
 (or equivalently of $B_z$)
 has a resonant peak  at $\omegarf=\Delta_z$ of width $2\gammadu$.
Since $\gammadu\geq 1/ T_2$, this width provides a lower bound on
 the intrinsic spin decoherence time $T_2$
 of a single dot-spin \cite{Spectroscopy}.
For weak tunneling, $\gO<2/T_2$, this bound saturates,
 i.e.\ the width $2\gammadu$ becomes $2/T_2$.
Further, we note that
 the $g$ factor of a single
 dot can be measured via
 the position of the peak
 in $I(\omegarf)$ [or in $I(B_z)$],
 which could provide a useful technique to study $g$ factor modulated
 materials \cite{QCReview,EffGFactor}.

{\it Pumping.} 
Next we consider the case of zero bias, 
 $\Delta\mu=0$, and $f_1=f_2$,
 but with
 $\gdl{l}\neq\gul{l}$.
Then, there is a finite current due to ``pumping'' \cite{PP}
 by the ESR source,
\begin{eqnarray}
\label{eqnCurrentDeltaZero}
&&I(\omegarf) = e \WRdu(\guO\gdT - \gdO\guT) f_1(\ESd) \Big/ \Big\{
  (\gdO \!+\! \gdT \!-\! \WRud)
\\ \nonumber &&
\quad  \times (\guO \!+\! \guT) f_1(\ESd)
 + (\WRud+\WRdu) (\guO+\gdO+\guT+\gdT)
\Big\},
\end{eqnarray}
where $\guO\gdT - \gdO\guT$
 determines the direction of the current.
As in
 Eqs.\ (\ref{eqnCurrentZeroT}) and (\ref{eqnCurrentLinResp}),
 Eq.\ (\ref{eqnCurrentDeltaZero})
 has resonant behavior
 and $T_2$ can also be measured.

In addition to setups
 using spin-polarized leads \cite{Recher} or spin-dependent tunneling,
 we now propose an alternative for producing $\gdl{l}\neq\gul{l}$.
A second dot [``dot 2'', see Fig.~\ref{figAcDot}(b)], acting as a spin filter,
 is coupled to the previous dot (``dot 1'') with tunneling amplitude $\tDD$.
The coupling of dot 2 to the lead shall be strong,
 leading to resonant tunneling with resonance width 
 $\Gamma_2=2\pi\nu|\tDL|^2$.
We require $\Gamma_2 < \ESu^2-\mu_2$ to
 neglect electron-hole excitations in lead 2 \cite{AndreevQD}.
We calculate the rates $\hgu$ and $\hgd$,
 for tunneling from dot 1 via dot 2 into lead 2
 in a $T$-matrix approach, with
 tunnel Hamiltonian $H_T = \HDD + \HDL$.
We evaluate the transition rates
$W_{fi} = 2\pi \biglb| \bra{f} 
  H_T \sum_{n=0}^\infty
  \left[ (\varepsilon_i + i\eta - H_0)^{-1} H_T \right]^n
  \ket{i} \bigrb|^2 \delta(\varepsilon_f-\varepsilon_i)$
 by summing up contributions from all orders in $\HDL$
 and taking $\eta\to+0$.
The Zeeman splitting $\Delta_z^2$ in dot 2
 shall be such that  $\Delta_z^1\not\approx\Delta_z^2$ \cite{ZeemanDotTwo}
 and $\Delta_z^2 > \ESd^1-\mu_2$.
This ensures by energy conservation
 that dot 2 is always in state $\spup$ after an electron has passed.
We now integrate over the final states in lead 2
 and obtain the  Breit-Wigner transition rate
 of an electron with spin down
 to go from dot 1 to lead 2 via the resonant level $\ES^2$ of dot 2,
\begin{equation}
\hgd = 
   \Gamma_2 |\tDD|^2  \big/
   \big[ (\ESu^1 - \ESu^2\big)^2 + \big(\Gamma_2/2\big)^2 \big].
\end{equation}
Since dot 2 is always in state $\spup$,
 tunneling of a spin $\uparrow$ 
 would involve the triplet level $\ETp$ on dot 2,
 and thus $\hgu$ is suppressed to zero
 (up to cotunneling contributions \cite{CT}).
The proposed setup is thus again described by
 Eqs.\ (\ref{eqnRhoDotuu})--(\ref{eqnRhoDotSd})
 with the tunneling rates $\WSdl{2}=\WdSl{2}=\WSul{2}=0$ 
 and $\WuSl{2}=\hgd$,
 and we can use all previous results (for one dot),
 but with $\gdT\to\hat{\gamma}^\downarrow$, $\guT\to0$,
 and $f_2(\ESu)=0$.
In particular, we see from Eq.\ (\ref{eqnCurrentDeltaZero})
 that for zero bias $\Delta\mu=0$ a current flows 
 from lead 1 via the dots 1 and 2 to lead 2.
We emphasize that this setup [see Fig. \ref{figAcDot}(b)]
 acts as a {\it ``spin inverter''}
 with spin up electrons as input
 and spin down electrons as output
 (and no transmission of spin down electrons).

{\it Spin read-out.}
We analyze now the time-dynamics of the read-out of a dot-spin
 via spin-polarized currents.
For this, we consider a dot coupled to fully spin polarized leads,
 such that $\Delta_z^{\rm leads}>\efermi>\Delta_z$ \cite{Recher}.
Since no electron with spin down 
 can be provided or taken by the leads (since $\nu_\downarrow=0$),
 the rates $\WSu=\WuS$ vanish
 (in contrast to the energy blocking $\WSu=0$ described above).
Thus, a current can only flow if initially the state on the dot is 
 $\spdown$\cite{PolarizedResonance},
 which allows to detect the initial spin state of the dot
 (strong measurement).
The goal is now to characterize a measurement time $\tmeas$ for
 the spin read-out.
For this,
 we need to keep track of the number of
 electrons $q$ which have accumulated in lead 2
 since $t=0$ \cite{Jong}
 (above we have only studied averaged currents),
 i.e.\ we now consider the states $\ket{n}\to\ket{n,\,q}$.
The time evolution of $\rho_D(q,\,t)$ (now charge-dependent)
 is described by
 Eqs.\ (\ref{eqnRhoDotuu})--(\ref{eqnRhoDotSd}),
 but with
 replacements
 $\WdSl{2}\,\rhoSS(q)\to\WdSl{2}\,\rhoSS(q-1)$ in Eq.\ (\ref{eqnRhoDotdd}) and
 $\WSdl{2}\,\rhodd(q)\to\WSdl{2}\,\rhodd(q+1)$ in Eq.\ (\ref{eqnRhoDotSS}).
Next, we consider the distribution function
 $P_i(q,\,t) = \sum_n \rho_n(q,\,t)$
 that $q$ charges have accumulated 
 in lead 2 after time $t$ 
 when the dot was in state $\ket{i}$ at $t=0$.
For a meaningful measurement of the dot-spin,
 the spin flip times $\Wud^{-1}$, $\Wdu^{-1}$, and $1/\HrfConst$ must 
 be smaller than $\tmeas$.
Eqs.\ (\ref{eqnRhoDotuu})--(\ref{eqnRhoDotSd})
 then decouple except 
 Eqs.\ (\ref{eqnRhoDotdd}) and (\ref{eqnRhoDotSS}),
which we solve for $\rhouu=1$ and $\rhodd=1$ at $t=0$
 (the general solution follows by superposition).
First, 
 if we start in state $\spup$, no charges tunnel through the dot
 and thus $P_\uparrow(q,\,t)=\delta_{q0}$.
Second,
 for the initial state $\spdown$,
 we consider $kT<\Delta\mu$
 and equal rates $\WSdl{1}=\WdSl{2}=W$.
We relabel the density matrix
 $\rhodd(q)\to\rho_{m=2q}$ and $\rhoSS(q)\to\rho_{m=2q+1}$,
 and Eqs.\ (\ref{eqnRhoDotdd}) and (\ref{eqnRhoDotSS}) become
 $\dot{\rho}_m = W (\rho_{m-1} - \rho_m)$,
 with solution
 $\rho_m(t) = (Wt)^m e^{-Wt} / m!$ (Poissonian distribution).
$P_i(q,\,t)$ then becomes
\begin{equation}
P_\downarrow(q,\,t) = \frac{(Wt)^{2q} e^{-Wt}}{(2q)!} 
 \left(1+\frac{Wt}{2q+1}\right).
\end{equation}
Experimentally, $P_\downarrow(q,\,t)$ can be determined
 by time series measurements.
The (inverse) signal-to-noise ratio is
 defined as the Fano factor \cite{BB,Devoret},
 which we calculate as
$F_\downarrow(t)=\expect{\delta q(t)^2}/\expect{q(t)} = 
 1/2 +   \big[3 - 2e^{-2Wt}(4Wt+1) - e^{-4Wt} \big] \big/
         4 \big(2Wt-1 + e^{-2Wt}\big)$,
 with $F_\downarrow$ decreasing monotonically
 from $F_\downarrow(0)=1$ to $F_\downarrow(t\to\infty)=1/2$ \cite{FanoCT}.
We can now quantify the measurement efficiency.
If, after time $\tmeas$, some charges $q>0$ have tunneled through the dot,
 the initial state of the dot was $\spdown$
 with probability 1
 (assuming that single charges can be detected via an SET \cite{Devoret}).
However, if no charges were detected ($q=0$),
 the initial state of the spin memory was 
 $\spup$ with probability 
 $1-P_\downarrow(0,\,t) = 
  1-(\WSdl{1}e^{-\WdSl{2}t}-\WdSl{2}e^{-\WSdl{1}t})/(\WSdl{1}-\WdSl{2})$,
 which reduces to $1-e^{-Wt}(1+Wt)$,
 for equal rates. 
Thus, roughly speaking, we find that 
 $\tmeas \gtrsim 2W^{-1}$, as expected.

{\it Rabi oscillations and Zeno effect.}
We show that coherent oscillations
 of the dot-spin induced by ESR
 lead to coherent oscillation in the current,
 again for spin polarized leads.
For $\mu_1 > \ESd > \mu_2$
 and $kT<\Delta\mu$,
 the current in lead 1 is
 $I_1(t) = -e \WSdl{1} \rhodd(t)$
 and $I_2(t) = e \WdSl{2} \rhoSS(t)$ in lead 2.
(Note that in general $I_1(t) + I_2(t) \neq 0$,
 since charge can accumulate on the dot.)
Thus, the time-dependence of $\rhodd$ and $\rhoSS$ 
 can be measured via the currents $I_{1,\,2}$.
Note that the spin-polarized electrons from lead 1 perform
 a projective measurement,
 leaving the dot-spin in either up or down state.
Thus, to obtain $I_{1,\,2}$ experimentally,
 an ensemble average is required, e.g\ by
 using an array of (independent) dots arranged in parallel
 or by time-series measurement over a single dot.
In Fig. \ref{figRabiOsc} we plot the numerical solution of
 Eqs.\ (\ref{eqnRhoDotuu})--(\ref{eqnRhoDotSd}),
 showing coherent Rabi oscillations of $\rhouu$, $\rhodd$
 and their decay to the stationary solution,
 dominated by the 
 spin decoherence $\gammadu$.
Thus, $\gammadu$ (and $1/T_2$)
 can be accessed here directly in the time domain \cite{PulsedESR}.

\begin{figure} 
\centerline{\psfig{file=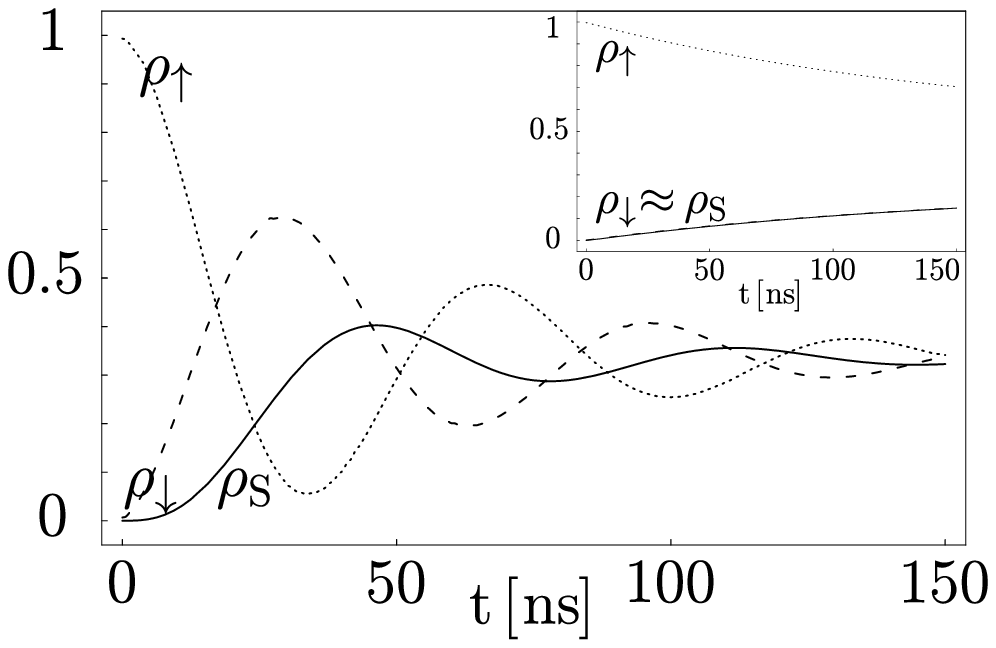,width=6cm}}
\caption[Rabi oscillations]{
\label{figRabiOsc}
Rabi oscillations visible
 in the time evolution of the density matrix
 $\rhouu$ (dotted), $\rhodd$ (dashed) and $\rhoSS$ (full line)
 for $\WSd=\WdS=4\times10^7 \: \rm{s}^{-1}$,  
 $T_1=1\: \mu{\rm s}$, 
 $T_2 = 300\: \rm{ns}$, and 
 $\HrfConst = 5\WSd$ 
 (corresponding to $B_x=10\: \rm{G}$ for $g=2$).
During the time span shown here, on average 3 electrons 
 have tunneled through the dot.
Here the spin decoherence 
 is dominated by the measurement process, $\WSd\gg1/T_2$,
 however, for weaker measurement,
 it will be determined by $T_2$.
In the inset we show the case of
 a strong measurement, $\WSd=\WdS=10^9 \: \rm{s}^{-1}$.
As a consequence of the Zeno effect (see text),
 the Rabi oscillations are suppressed.
Further, $\rhodd$ and $\rhoSS$ are indistinguishable since
 $\spdown$ and $\ket{S}$ equilibrate rapidly due to the increased tunneling.
}
\end{figure}

Finally, increasing $\WSd$, the coherent oscillations 
 of $\rhouu$, $\rhodd$ become suppressed (see Fig.~\ref{figRabiOsc})
 due to increased transfer of charges which perform
 a continuous strong measurement on the dot-spin.
This suppression, known as Zeno effect \cite{Peres},
 occurs in $\rhouu$, $\rhodd$ and thus is observable
 in the input current $I_1(t)$.

{\it Conclusions.}
We have proposed a setup to measure the
 single spin decoherence time $T_2$ 
 of a dot in Coulomb blockade regime, coupled to leads,
 via the stationary and time-dependent current
 by using ESR techniques. We have discussed pumping and read-out
processes.

{\it Acknowledgements.}
We thank 
 G. Burkard, F. Marquardt, P. Recher, and E. Sukhorukov 
 for discussions.
This work has been supported by the Swiss NSF.

\ifpreprintsty\else\newpage\end{multicols}\fi


\begin{thebibliography}{}


\bibitem[*]{EmailEngel}       Email: Hans-A.Engel@unibas.ch
\bibitem[\mbox{$\dagger$}]{EmailLoss}        Email: Daniel.Loss@unibas.ch


\bibitem{Kikkawa}
D.D. Awschalom, J.M. Kikkawa, Phys.\ Today {\bf 52}(6), 33 (1999).


\bibitem{Gupta}
J.A. Gupta\etal{
 D.D. Awschalom, X. Peng, and A.P. Alivisatos},
 Phys. Rev. B {\bf 59}, R10421 (1999).


\bibitem{Roukes}
F.G. Monzon, M.L. Roukes, 
 J. Magn. Magn. Mater. {\bf 198}, 632 (1999).


\bibitem{Fiederling}
R. Fiederling\etal{
 M. Keim, G. Reuscher, W. Ossau, G. Schmidt,  A. Waag,
 and L.W.  Molenkamp},
Nature {\bf 402}, 787 (1999);
Y. Ohno\etal{
 D.K. Young, B. Beschoten, F. Matsukura, H. Ohno,
 and D.D. Awschalom},
{\it ibid.} {\bf 402}, 790 (1999).


\bibitem{Ensslin}
S. L\"uscher\etal{
 T. Heinzel, K. Ensslin, W. Wegscheider,
 and M. Bichler},
cond-mat/0002226.


\bibitem{Fujisawa}
T.~Fujisawa, Y.~Tokura, Y.~Hirayama, cond-mat/0010437.


\bibitem{Loss97}
D. Loss, D.P. DiVincenzo,
Phys.\ Rev.\ A {\bf 57}, 120 (1998).


\bibitem{QCReview}
G. Burkard, H.-A. Engel, D. Loss,
 Fortschr. Phys. {\bf 48}, 965 (2000).

\bibitem{KhaetskiiNazarov}
A.V. Khaetskii, Y.V. Nazarov,
 cond-mat/0003513.


\bibitem{Schoen}
For measurement process in Josephson qubits see
Y. Makhlin, G. Sch\"on, A. Shnirman,
 cond-mat/0001423.


\bibitem{kouwenhoven}
L.P. Kouwenhoven, G. Sch\"on, L. L. Sohn, 
 in Mesoscopic Electron Transport, 
 NATO ASI Series E, Vol. {\bf 345}
 (Kluwer Academic Publishers, Dordrecht, 1997). 

\bibitem{ZeemanDotTwo}
We assume that dot 2 remains unaffected by the ESR field,
 achieved e.g.\ by applying $B_x$ and/or $B_z$ locally
 or with different $g$ factors;
 to account for this,
 we simply assume $\Delta_z^1\not\approx\Delta_z^2$.


\bibitem{taruchaKouwenhoven}
S. Tarucha\etal{
 D.G. Austing, T. Honda,
 R.J. van der Hage and L.P. Kouwenhoven},
Phys. Rev. Lett. {\bf 77}, 3613 (1996).

\bibitem{Blum}
K. Blum, {\it Density Matrix Theory and Applications}
 (Plenum Press, New York, 1996), Chap.\ 8.

\bibitem{Details}
H.-A. Engel, D. Loss, 
 unpublished.


\bibitem{CT}
We work in the sequential tunneling regime with
 negligible cotunneling contributions \cite{kouwenhoven,Recher,NoiseLong}.

\bibitem{Spectroscopy}
Note the similarity to spectroscopy,
 where absorption or emission line widths
 provide information on decoherence.


\bibitem{EffGFactor}
The $g$ factor in such materials can be controlled by shifting
 the equilibrium position of the electrons in the dot
 from one layer to another by electrical gating \cite{QCReview}.
This allows local manipulation of electron spins, 
 being essential in concepts for quantum computers 
 \cite{Loss97,QCReview}.


\bibitem{PP}
For non-magnetic photon-assisted pumps see e.g.\ \cite{kouwenhoven}.


\bibitem{Recher}
P. Recher, E.V. Sukhorukov, D. Loss,
 Phys. Rev. Lett. {\bf 85}, 1962 (2000).


\bibitem{AndreevQD}
P. Recher, E.V. Sukhorukov, D. Loss,
 cond-mat/0009452.


\bibitem{PolarizedResonance}
In the stationary regime and for $\Delta_z>kT$,
 the current becomes blocked due to
 spin relaxation ($\Wud$).
However, this blocking can be removed
 by the ESR field producing spin-flips on the dot
 (with rate $\Womegarf$).
This competition leads, for $\Womegarf<\Wud$,
 again to a  stationary current with resonant structure,
$I(\omegarf)=e\,(\gO\gT\,\WRdu)/[\gT\,\Wud + (\gO+\gT)\,\Wdu]$,
 from which $\gammadu$ (and 1/$T_2$) can be measured.


\bibitem{Jong}
M.J.M. Jong,
 Phys. Rev. B {\bf 54}, 8144 (1996).


\bibitem{BB}
Ya.M. Blanter, M. B\"uttiker,
 Phys. Rep. {\bf 336}, 1 (2000).

\bibitem{Devoret}
M.H. Devoret, R.J. Schoelkopf,
 Nature {\bf 406}, 1039 (2000).
    
\bibitem{FanoCT}
Note that for dot-spin $\spup$,
 only weak cotunneling occurs
 with Fano factor $F_\uparrow=1$ \cite{NoiseLong}.


\bibitem{PulsedESR}
An alternative method for measuring
 $T_2$ is pulsed ESR.
An initial spin state, say, $\spdown$,
 is rotated with a $\pi/2$ pulse into the $xy$-plane
 and starts to precess with frequency $\Delta_z$.
After a delay $\tau$,
 the (remaining) coherent part of the spin
 is rotated by another $\pi/2$ pulse 
 and finally $\rhodd$ is measured via the current.
Since the final spin direction,
 and thus $\rhodd$,
 depends on the precession angle,
 the measured current
 is an oscillating function of $\tau$ with frequency $\Delta_z$,
 which decays on a timescale $T_2$.

\bibitem{Peres}
A. Peres, {\it Quantum Theory} 
 (Kluwer, Amsterdam, 1993).

\bibitem{NoiseLong}
E.V. Sukhorukov, G. Burkard, D. Loss,
 cond-mat/00010458.


\end{thebibliography}
\end{document}